# FORECASTING THE FULL DISTRIBUTION OF EARTHQUAKE

# NUMBERS IS FAIR, ROBUST AND BETTER

Shyam Nandan[1], Guy Ouillon[2], Didier Sornette[3], Stefan Wiemer[1]

[1]ETH Zürich, Swiss Seismological Service, Sonneggstrasse 5, 8092 Zürich, Switzerland

[2]Lithophyse, 4 rue de l'Ancien Sénat, 06300 Nice, France

[3]ETH Zürich, Department of Management, Technology and Economics, Scheuchzerstrasse 7, 8092 Zürich, Switzerland

Corresponding author: Shyam Nandan (nandans@ethz.ch)

# Abstract:

Forecasting the full distribution of the number of earthquakes is revealed to be inherently superior to forecasting their mean. Forecasting the full distribution of earthquake numbers is also shown to yield robust projections in the presence of "surprise" large earthquakes, which in the past have strongly deteriorated the scores of existing models. We show this with pseudo-prospective experiments on synthetic as well as real data from the Advanced National Seismic System (ANSS) database for California, with earthquakes with magnitude larger than 2.95 that occurred between the period 1971-2016. Our results call in question the testing methodology of the Collaboratory for the study of earthquake predictability (CSEP), which amounts to assuming a Poisson distribution of earthquake numbers, which is known to be a poor representation of the heavy-tailed distribution of earthquake numbers. Using a spatially varying ETAS model, we demonstrate a remarkable stability of the forecasting performance, when using the full distribution of earthquake numbers for the forecasts, even in the presence of large earthquakes such as Mw 7.1 Hector Mine, Mw 7.2 El Mayor-Cucapah, Mw 6.6 Sam Simeon earthquakes, or in the presence of intense swarm activity in Northwest Nevada in 2014. While our results have been derived for ETAS type models, we propose that all earthquake forecasting models of any type should embrace the full distribution of earthquake numbers, such that their true forecasting potential is revealed.

# 1 Introduction

In recent years, significant efforts have been devoted towards devising and testing models that yield probabilities of earthquake occurrence within given space-time-magnitude windows [Schorlemmer and Gerstenberger, 2007; Field, 2007; Werner et al., 2010; Schorlemmer et al., 2010b; Zechar et al., 2010; Helmstetter et al., 2007; Werner et al., 2011; Helmstetter and Werner, 2014; Kagan and Jackson, 2011; Zechar et al., 2013; Rhoades et al., 2014; Stader et al., 2017]. A general framework for testing model forecasts has been developed within the international Collaboratory for the study of earthquake predictability (CSEP). As pointed out by Schorlemmer et al. [2018]: "The fundamental idea of CSEP is simple in principle but complex in practice: forecasting models should be tested against future observations to assess their performance, thereby ensuring an unbiased test of the forecasting power of a model."

Consider a model ($\Lambda$) that is currently tested in CSEP. It performs a forecast, taking the form of the expected number of earthquakes $\{\lambda(i), i = 1, \dots, n\}$ in $n$ predefined space-time-magnitude bins. Given the observations of the numbers of earthquakes $\Omega = \{\omega(i), i = 1, \dots, n\}$, in each of the $n$ predefined bins, the model $\Lambda$ obtains a log likelihood score $LL = \sum_{i=1}^{n} \omega(i) \ln \lambda(i) - \lambda(i) - \ln \omega(i)!$. Different models are then compared based on their log likelihood score $LL$. This formula of the log likelihood score $LL$ shows that this model evaluation scheme inherently assumes that the distribution of the number of earthquakes in each of the bins i follows a Poisson distribution $e^{-\lambda(i)} \frac{\lambda(i)^{\omega(i)}}{\omega(i)!}$. This choice is natural only when the forecast provides solely the mean rate since, by the maximum entropy principle, the Poissonian distribution is the most parsimonious, i.e. its choice amounts to assuming the minimum number of additional assumptions, given the forecast of the mean number. However, if the forecast provides more information than the mean number, then this choice of a Poisson distribution is not justified. Moreover, the Poisson distribution is unique in its property that $\lambda(i)$ is both the mean and the variance of the number of earthquakes in bin i. But, for skewed distributions as is well-known to occur in the distribution of earthquake numbers [Saichev and Sornette, 2006a; 2007], the mean value can be a very poor measure of the most probable number of earthquakes, the latter being, by definition, more likely to occur and thus more relevant for the forecast [Redner, 1990].  For instance, for the log-normal distribution with unit log-variance (resp. variance equal to 2, resp. 3), the mean is approximately 4.5 (resp. 20, resp. 90) times larger than the most probable value! Therefore, the models, which  predict distributions deviating from the Poisson distribution, as they should if they are to represent more correctly the highly skewed nature of the real distributions of earthquakes, are forced to be inconsistent with their own best judgement, leading to the violation of the maxim: "A requisite forecast should always correspond to the forecasters best judgement" [Murphy, 1993]. Since, these models are forced to not adhere to their best judgement, we argue that they are, by construction,  on a weaker footing than the models whose best judgement is in line with the Poissonian assumption. By extension, the testing metric that is used by CSEP to rank the models with respect to each other can be arguably called "improper", as it offers more advantage to those models that strictly or closely adhere to the Poissonian assumption. CSEP is aware of this limitation and plans to overcome it in the near future by allowing the modelers to specify not only the expected number of earthquakes in each bin, but also the full distribution of target earthquakes in each bin, as well as correlations between bins to account for epistemic uncertainties [Schorlemmer et al., 2018].

Here, we demonstrate the crucial importance of such a step, by revealing that the forced Poissonian assumption might have hindered several models (the ETAS model, in particular) from displaying their true forecasting potential relative to declustering-based models such as the ones proposed by Helmstetter et al. [2007], which adhere to the Poissonian assumption. More specifically, we conduct CSEP-style forecasting experiments to test if the performance of the ETAS model that forecasts the full distribution of the number of earthquakes in a given space-time-magnitude bin is significantly better than the variant of the same ETAS model that has been forced to use only its estimate of the mean rate, along with the forced assumption that the distribution of the number of earthquakes in a given bin is Poissonian. We first present the proof of concept on a synthetic catalogue, and then extend this analysis to the Californian earthquake catalogue, and show that the correctly scored ETAS model now systematically and appreciably outperforms the model of Helmstetter et al. [2007], even within space-time bins where both models used to yield comparable scores due to the improper Poissonian-based scoring.

## 2. The ETAS model

Several flavours of the ETAS model exist in the literature [Zhuang et al., 2002, 2004; Ogata, 1998]. In this paper, we use the one in which the conditional seismicity rate ($\lambda$) at any given location ($r$), time ($t$) and magnitude ($m$) depends on the history of seismicity $H_t$ that occurred before $t$ in the following way:

$$\lambda(r,t) = \mu + \sum_{t_i < t} K \, exp[a(m_i - M_0)] \times (t - t_i + c)^{-1-\omega} \exp(-\frac{t - t_i}{\tau}) \times$$
$$\{\|r - r_i\|^2 + d \, exp[\gamma m_i]\}^{-1-\rho} \tag{1}$$

In Equation 1:

1. $\mu$ quantifies the rate of background earthquakes.

2. $K$ and $a$ are the parameters of the productivity law, which quantifies the expected number of aftershocks (with magnitudes larger than $M_0$) triggered by an earthquake of magnitude $m_i$.

3. $(t - t_i + c)^{-1-\omega}\exp(-\frac{t-t_i}{\tau})$ is the time kernel, which quantifies the time distribution of aftershocks triggered by an earthquake that occurred that time $t_i$. The exponential taper, $\exp(-\frac{t-t_i}{\tau})$, ensures that the integral of the time kernel is finite even for negative values of $\omega$. In typical implementations of the ETAS model, $\omega$ is restricted to be positive in absence of such an exponential taper.

4. $\{\|r - r_i\|^2 + d \, exp[\gamma(m_i - M_0)]\}^{-1-\rho}$ is the spatial kernel, which quantifies the distribution of the location of aftershocks triggered by an earthquake that occurred at location $r_i$.

5. $M_0$ is the minimum magnitude of the earthquakes that can occur and is assumed to be equal to the magnitude of completeness ($M_c$) of the catalog. Note that $M_0$ is also assumed to be equal to the minimum magnitude of earthquakes that can trigger their own aftershocks.

6. Although not explicitly included in Equation 1, ETAS model assumes that the magnitudes ($m$) of both the background as well as the aftershocks are the distributed according to Gutenberg-Richter law: $f(m) \sim \beta e^{-\beta(m-M_0)}$.

# 3. Data

## 3.1 The synthetic ETAS catalogue

To conduct pseudo-prospective experiments with synthetic data, we use the ETAS model to generate synthetic earthquake catalogues. We generate 200 years long catalogues of earthquakes ($M \geq 1$) within a $1000 \times 1000\ km^2$ area (see Figure S1). The parameters that are used to generate this catalog are: $\mu = 9168$ earthquakes ($M \geq 1$) per year and square kilometer; $d = 0.12\ km^2$; $\rho = 0.6$; $\gamma = 1.2$; $\omega = 0.1$; $c = 6.7 \times 10^{-3} days$; $\tau = 27.38\ years$; $K = 1.35 \times 10^{-4}$ and $a = 2.9$. The magnitudes of the earthquakes are generated assuming a Gutenberg-Richter (GR) law with $\beta = 2.3$, which is equivalent to the typically reported exponent of GR law in base 10 ($b = 1$). The parameter combination used to generate the synthetic catalog correspond to a branching ratio of ~0.96, i.e. each earthquake on an average produces 0.96 offspring. The values of these parameters are based on the typical estimates of these parameters for Californian catalog. However, it is important to note that our results that are based on the synthetic catalog can be showed to be robust with respect to the choice wide range of parameter combinations.

## 3.2 Californian catalogue

We also conduct pseudo-prospective experiments with earthquakes that occurred in and around the state of California (Figure S2). The catalogue is obtained from the Advanced National Seismic System (ANSS) database (see Data and Resources). We only use the earthquakes with magnitude larger than 2.95 that occurred between the period 1971-2016 in the spatial polygon defined by Schorlemmer et al. [2007]. The competing forecasting models use the catalog between 1981 and 2016 as the primary catalog, while the earthquakes that occurred between 1971-1981 are treated as part of the auxiliary catalog. Both primary and auxiliary catalogs are used to calibrate the models competing in the experiments; however, the earthquakes in the auxiliary catalog are restricted to act only as parents of the earthquakes in the primary catalog. The consideration of such an auxiliary catalog is relevant for the calibration of ETAS-based models (see Wang et al. [2010], for instance), to avoid the situation that earthquakes in the early part of the primary catalog might systematically appear as orphans (i.e. background events) due to the artificial lack of any ancestor [Sornette and Werner, 2005; Saichev and Sornette, 2006b].

# 4. Method

## 4.1 Test set up

The area encompassing the catalog is divided into $N$ cells with dimension $x^2$. Any model is evaluated based on the probabilities that it assigns to the observed numbers of earthquakes ($M \geq M_t$) in all the N spatial bins during the testing period, where $M_t$ is the lower magnitude threshold of the testing earthquake catalog. More specifically, if a model assigns the probability $(Pr(n_k))$ to the occurrence of $n_k$ earthquakes in the $k^{th}$ bin, the log likelihood score that such a model obtains is given by:

$$LL = \sum_{k=1}^{N} \ln[\Pr(n_k)] \tag{2}$$

If the model only specifies the expected number of earthquakes, $\lambda_k$, and assumes that the distribution of the number of earthquakes in the bin is Poissonian, then the log likelihood score of such a model obtains is given by:

$$LL = \sum_{k=1}^{N} n_k \ln \lambda_k - \lambda_k - \ln(n_k!) \tag{3}$$

As mentioned in the introduction, the Poisson distribution is justified only when the forecast provides solely the mean number of earthquakes in a given bin, as a consequence of the maximum entropy principle.

To construct the forecast, the model can only use the data in the training period. Using this "allowed" training data, the model makes the forecast over the next T years that immediately follows the training period. Our experiments with the synthetic data start with the first training period covering the first 30 years of the catalogue, which is used to make the forecasts for the testing period (30, 30+T). The training period is then updated by T years to [0, 30+T], which is then used to make a forecast for the testing period (30+T, 30+2T). This process of updating the training and testing periods is repeated for 20 different testing periods. In order to check for the robustness of the results, in the case of synthetic catalogs, we try different combinations of:

(i)     spatial resolution, x, at which the forecast is issued: $x \in \{10, 50, 100, 500, 1000\} km$

(ii)   magnitude threshold, $M_t$, of the testing catalog: $M_t \in \{1, 2, 3, 4, 5\}$

(iii)  duration of the testing period, T: $T \in \{1, 5\}$ years

In summary, we conduct $20 \times 2 \times 5 \times 5 = 1000$ experiments with different combinations of the training period, duration of the training period, minimum magnitude threshold of the target catalogue and spatial resolution at which the forecast is specified.

The setup of the experiments with the real dataset is similar to the synthetic counterpart. Specifically, we conduct five pseudo-prospective experiments with multiple training and testing periods. In each of these experiments, the earthquakes in the training period are used to calibrate the models, which forecast the seismicity during a T year long testing period that immediately follows the training period. The competing models specify their forecast in the form of the full distribution of the number of earthquakes with magnitude larger than $M_t$ in each of the $0.1° X 0.1°$ grid cells into which the testing region is divided. To be consistent with the CSEP experiments, we use the same definition of the testing region as well as the same testing grid (7682 cells in total) as used by Zechar et al. [2013]. The five experiments correspond to different combinations of the duration (1 or 5 years) and of the minimum magnitude threshold ($M_t = 2.95, 3.95$ or $4.95$) of the testing catalog. In Table 1, we give the summary of all the experiments. Note that we do not conduct one-year long experiments with $M_t = 4.95$ as there is not enough earthquakes in many of the testing periods to compute reliable statistics.

## 4.2 Competing Models

With both real and synthetic data, we conduct the horse-race between the two "philosophically" different variants of the same ETAS model. The first variant (Model 1) forecasts the full distribution of the number of earthquakes in each of the spatial bins, which it obtains by conducting many simulations for the testing period using the training catalogue as initial conditions, and parameters that has been estimated during the training period. The second variant (Model 2) only forecasts the expected number of earthquakes in each of the spatial bins, which is equal to the average number of earthquakes observed in all the simulations combined in a given spatial bin. This model further assumes that the distribution of the number of earthquakes in each of the bins is Poissonian.

In the case of synthetic data, the base ETAS model that the two variants depend on is the one that has been used to generate the "true" catalogue itself. Normally, we would have to estimate the parameters of the model by calibrating it on the training data. However, as we are dealing with the "best-case" scenario, we assume that we can estimate the best possible parameter estimates (i.e. the true parameters themselves) using the data in the training catalogue. This assumption enormously simplifies our task, as we can simply use the true parameters values as if they were estimated for any training catalogue. On the other hand, as the generating parameters of the real catalog are unknown a priori, we must estimate them for each of the training periods. Furthermore, given the heterogeneity of the Earth's crust, it is likely that the parameters of the ETAS model would also exhibit spatial variability. Using an objective method to invert the spatial variation of ETAS parameters (See Supplementary Text S1), Nandan et al. [2017] found that the parameters of the ETAS model indeed show substantial spatial variation. Building on the results of Nandan et al. [2017], we will use the ETAS model with spatially variable parameters (SVETAS) for the pseudo-prospective experiments with the Californian catalogue and study how the assumption of a Poisson distribution of earthquake numbers impact its performance.

Using the estimated parameters and the training catalogue, we simulate numerous (~5,000,000) catalogs for a given testing period. We provide the details of the simulation algorithm in the Supplementary Text S2.

Using these simulated catalogues, we compute the empirical distribution of the number of earthquakes in each of the spatial bins. The mean rate in each of the bins, which is the forecast of Model 2, can be easily estimated by dividing the total number of earthquakes in each bin by the total number of simulations conducted.

While we could directly use the distribution of the number of earthquakes in each of the bins estimated from the simulations as providing the forecast of Model 1, this is not a wise choice because this estimated distribution has finite support as a result of the finite number (despite being large ~5,000,000) of generated catalogs. Furthermore, the empirical distribution is imperfectly sampled and, due to statistical fluctuations always present in any finite size simulation, it can have holes within, i.e., not all values of n (number of earthquakes observed in a simulation) within the finite support are sampled during the simulations. Those values of n thus have an estimated probability of occurrence being 0. If one then would use this estimated distribution to forecast, it is possible that the observed number of earthquakes in a bin during the testing period either could fall in one of the holes of the estimated distribution or even beyond its finite support. As a result, the log likelihood score of the model becomes $-\infty$ and is surely rejected as the worst model. To avoid such unfortunate scenarios that would amplify unavoidable statistical fluctuations of the construction of the distribution of earthquake numbers, we smooth the empirical distribution in each of the spatial bins to obtain the estimates of the probabilities for unobserved values of n. The details of the smoothing method are given in Supplementary Text S3.

Figure 1 (panels a to e) shows the estimated distribution (blue crosses) of the number of earthquakes ($M \geq 5$) obtained from the simulated catalogs for the testing period (30, 35) in a randomly selected spatial bin in the experiment with the synthetic data. The five panels show the empirical distribution for different choices of the spatial bin resolution ($1000 \times 1000\ km^2, 10 \times 10\ km^2, 50 \times 50\ km^2, 100 \times 100\ km^2, 500 \times 500\ km^2$). As predicted from the ETAS model and already reported on real empirical data [Saichev and Sornette, 2006a; 2007; Saichev et al., 2005], these estimated distributions clearly exhibit heavy tails and the Poisson distribution, with its thin tail, is an extremely poor choice for their description as seen from the comparison of the estimated distributions with the dashed black curves. These Poisson distributions are obtained by using the estimated mean, $\lambda$ (indicated at the top of each panel), equal to the average number of earthquakes observed in all the conducted simulations. In these panels, we also show the smoothed versions of the estimated distribution (red curve), obtained using the smoothing method described in Supplementary Text S3. One can check that the smoothed pdfs describe the empirical distributions very well.

# 5.Results and Discussion

## 5.1 Performance of Model 1 vs Model 2 on synthetic data

To evaluate the performance of Models 1 and 2, we use the log likelihood metrics defined in equations (2) and (3), respectively. Using the overall likelihood score of the two models, we then compute a new metric: the Mean Information gain (MIG), which is simply the difference in the Log Likelihoods of the two models divided by the total number of earthquakes that are observed during the testing period. Figure 2 shows the time series of MIG of Model 1 over Model 2, which is obtained from 20 five-years long experiments. In this figure, we report the time series for the five choices of magnitude threshold $M_t$ for the testing catalog, and the 5 choices of the spatial bin resolution. The black stars show the number of earthquakes that were observed during each of the testing periods. A similar plot as Figure 2 for 20 one-year long testing periods is shown in Figure S4. We find that Model 1 achieves substantial positive information gains over Model 2 in nearly all the 20 time periods considered in both five-year and one-year long experiments. Furthermore, this result is robust with respect to the choice of the magnitude threshold of the target catalog as well as the resolution of the spatial bins with which the models were evaluated. Note that the information gain of Model 1 over Model 2 systematically increases with increasing spatial resolution with which the models are evaluated. One can observe that there is a systematic positive correlation between the information gain of Model 1 over Model 2 and the occurrence of a large number of earthquakes during the testing period. This positive correlation is obvious for the small magnitude thresholds ($M_t = 1\ to\ 4$) but is less apparent for the largest magnitude threshold of the target catalog. This apparent lack of correlation could be attributed to the statistical fluctuations in the estimate of the MIG due to an insufficient number of earthquakes ($M \geq 5$) occurring during some testing periods.

## 5.2 Information gain of SVETAS 1 over SVETAS 2 in experiments with Californian catalogue

Figure 3 shows the mean information gain (MIG) that the two models (SVETAS 1 and SVETAS 2) obtain over a common null model during different testing periods using the red and blue circles respectively. The null model that we have used is the spatially and temporally homogeneous Poisson process (STHPP hereafter). The STHPP model forecasts the same rate of earthquakes in each of the testing grid cells, which is equal to $\frac{N(\geq M_t)}{N_{grid}} \times \frac{T_{test}}{T_{train}}$, where $N(\geq M_t)$ is the number of earthquakes with magnitude larger than the predefined magnitude threshold $M_t$ of the testing catalog that have been observed during the training period within the testing polygon; $N_{grid}$ is the number of $0.1° X 0.1°$ grid cells that divide the testing polygon; $T_{test}$ and $T_{train}$, respectively, are the duration of the testing and the training periods. The red and blue shaded region represent the 95% confidence interval of, respectively, the MIG of SVETAS 1 and SVETAS 2 models. The magnitude threshold of the testing catalog as well as the time duration of the testing period are indicated at the top of each panel.

In these figures, we also show the MIG obtained by the smoothed seismicity model, which is based on the declustering algorithm proposed by Reasenberg [1985]. We call this model D-HKJ after the initial of the authors [Helmstetter et al., 2007] who first used it during the first RELM experiment. In order to construct this forecast, the training catalog is first declustered using the Reasenberg's algorithm, whose parameters are set to fixed values as prescribed by Helmstetter et al. [2007]. The outcome of applying this algorithm on the training catalog is the "hard" classification of earthquakes in to two groups: independent earthquakes and aftershocks, with independence probability IP equal to 1 and 0 respectively. The locations

of all earthquakes are then smoothed using isotropic Gaussian kernels weighted according to

the independence probabilities (1 or 0 in this case) of the earthquakes. In a given spatial bin

(pre-defined by CSEP) with index I and spatial extent $S_I$, the smoothed estimate of the number

$\widehat{N}_{bkg}$ of background earthquakes during the training period is given by:

$$\widehat{N}_{bkg} = \sum_i \int_{S_I} k_i(x, y) \, dx \, dy$$
$$k_i(x, y) = \frac{IP_i}{2\pi\sigma_i^2} \exp\left[-\frac{1}{2}\left\{\frac{(x - x_i)^2 + (y - y_i)^2}{\sigma_i^2}\right\}\right]$$

(5)

where $k_i(x, y)$ is a 2D isotropic Gaussian kernel with bandwidth $\sigma_i$, weighted by the independence probability, $IP_i$, of the $i^{th}$ earthquake in the training catalogue. In Equation (5), $x_i$ and $y_i$ represent the location of the $i^{th}$ earthquake in the catalogue; $\sigma_i$ is taken as the distance

of the nearest neighbour background earthquake from the $i^{th}$ earthquake. Finally, the forecast

of the total seismicity rate during the testing period according to the model D-HKJ is

constructed by first normalising the spatial density of the background earthquakes such that it

integrates to 1 over the predefined testing polygon, then by upscaling the spatial pdf by a factor

$\frac{T_{test}}{T_{train}} N(\geq M_t)$, where $N(\geq M_t)$ is the number of earthquakes (both mainshocks and

aftershocks) with magnitude larger than the predefined magnitude threshold $M_t$ of the testing catalogue that have been observed during the training period within the testing polygon. This upscaling is a necessary inconsistency as, in the original state, the smoothed seismicity model would only forecast the rate of background earthquakes during the testing period, which is just a minor part of the seismicity that is going to occur during that period. It is important to note that the D-HKJ model, by its construction, assumes that the rate of future earthquakes in each of the bins is described by a simple rescaling of the future background seismicity rate. It thus has to be considered as a genuine Poissonian model, and it should be tested as such.

We find that the MIG of the SVETAS 2 is nearly always higher than that of D-HKJ. However, there are a few exceptions. For instance, in the 1 year long experiments with magnitude thresholds 3.95 and 2.95, the D-HKJ model either slightly outperforms or performs nearly as well as the SVETAS 2 model during years 2000, 2001 and 2014. This means that, when judged in terms of average numbers, i.e., assuming a Poisson distribution of earthquake numbers, the forecasts of the spatially varying ETAS model are in general clearly superior, but can be sometimes inferior to the intrinsic Poisson-based D-HKJ model. Comparing the MIG of the SVETAS 1 model with that of the D-HKJ model reveals that the Poissonian assumption is at the origin of these results. Indeed, the SVETAS 1 model constantly and significantly outperforms not only D-HKJ, but also SVETAS 2 during all testing periods and for all magnitude thresholds of the testing catalogue. This result demonstrates that, when the forecasts of the SVETAS model are forced to assume a Poissonian distribution (i.e. when using the SVETAS1 model), it yields a not-as-good performance relative to the D-HKJ model as it should, given that the D-HKJ model is built on the inherent assumption of a Poissonian distribution. On the other hand, the SVETAS 1 model, which accounts for the true distribution of forecasted rates, performs much better. This comparison demonstrates the flaw in the CSEP's current testing methodology, which tends to favor models with inherent Poissonian assumption.

It is also interesting to note that, during special testing periods such as during the year 1999 and 2010, when the Mw 7.1 Hector Mine and Mw 7.2 El Mayor-Cucapah earthquakes occurred, the SVETAS 1 model obtains exceptionally high MIGs of ~4.5 and ~5.2, respectively. D-HKJ and SVETAS 2 models, in comparison, show relatively mediocre performances during these periods. Similarly, when the Mw 6.6 Sam Simeon earthquake struck in 2003 followed by its intense aftershock sequence, or when an intense swarm activity occurred in Northwest Nevada in 2014, the D-HKJ as well as SVETAS 2 models exhibit deteriorated performances. In contrast, the SVETAS 1 model held its ground and showed consistent performance during all these periods. These examples demonstrate that, by accounting for its true forecast distribution, the SVETAS 1 model becomes much more resilient to surprises without incorporating ad-hoc concepts such as a minimum "water level" forecast [Wiemer and Schorlemmer, 2007; Hiemer and Kamer 2016; Kagan and Jackson, 1994, 2000, 2012].

## 6 Conclusion

With pseudo-prospective experiments on synthetic as well as real data, we have demonstrated a very significant information gain of the forecasts of earthquake numbers within space-time-magnitude bins when using the full projected distribution of earthquake numbers, as opposed to just forecasting the mean numbers of earthquakes as done in the current CSEP testing methodology. Indeed, the CSEP testing methodology amounts to assuming a Poisson distribution of earthquake numbers, which is known to be a very poor representation of the heavy-tailed distribution of earthquake numbers. First, this Poissonian assumption tends to favour models that adhere to it, with the consequence of a possible misclassification of competing models. Second, by accounting for the true heavy tailed distributions of the forecasted earthquake numbers, we showed that the earthquake forecasting models can become resilient to the occurrence of "surprise" large earthquakes, which in the past have strongly deteriorated the scores of existing models. Using a spatially varying ETAS model, we have demonstrated a remarkable stability of the forecasting performance, when using the full distribution of earthquake numbers for the forecasts, even in the presence of large earthquakes such as Mw 7.1 Hector Mine, Mw 7.2 El Mayor-Cucapah, Mw 6.6 Sam Simeon earthquakes, or in the preence of intense swarm activity in Northwest Nevada in 2014.

While our results have been derived for ETAS type models, the idea of embracing the full distribution of earthquake numbers should be extended to earthquake forecasting models of other types, such that their true forecasting potential is revealed. The concept of using the full distribution rather than the mean value to represent and characterize complex physical systems actually dates back to the seminar article by Nobel Prize winner P.W. Anderson [1958], whose insight opened the road towards replacing averages by the full distribution in all scientific fields of investigations. It is thus more than ripe time for CSEP and the statistical seismological community to embrace this approach, further motivated by our very encouraging results.

## Data and Resources

The dataset used in this study can be obtained from the website.

http://www.quake.geo.berkeley.edu/anss/catalog-search.html.

## Acknowledgement

We wish to thank Y. Kamer to provide valuable feedbacks during the development of the work.

## Mailing Address

Shyam Nandan: nandans@ethz.ch

Guy Ouillon: Ouillon@aol.com

Didier Sornette: dsornette@ethz.ch

Stefan Wiemer: stefan.wiemer@sed.ethz.ch

# Tables

Table 1: Description of the pseudo-prospective experiments conducted with the Californian catalogue.

| Experiment | Duration of testing period (years) | Magnitude threshold of the testing catalogue | Description |
|---|---|---|---|
| 1 | 5 | 4.95 | Considers testing windows [1999:2003], [2000:2004], etc, until [2012:2016]. |
| 2 | 5 | 3.95 | Same as Experiment 1 |
| 3 | 5 | 2.95 | Same as Experiment 1 |
| 4 | 1 | 3.95 | Considers testing windows [1999:2000), [2000:2001), etc, until [2016:2017). |
| 5 | 1 | 2.95 | Same as Experiment 4 |

# Figures

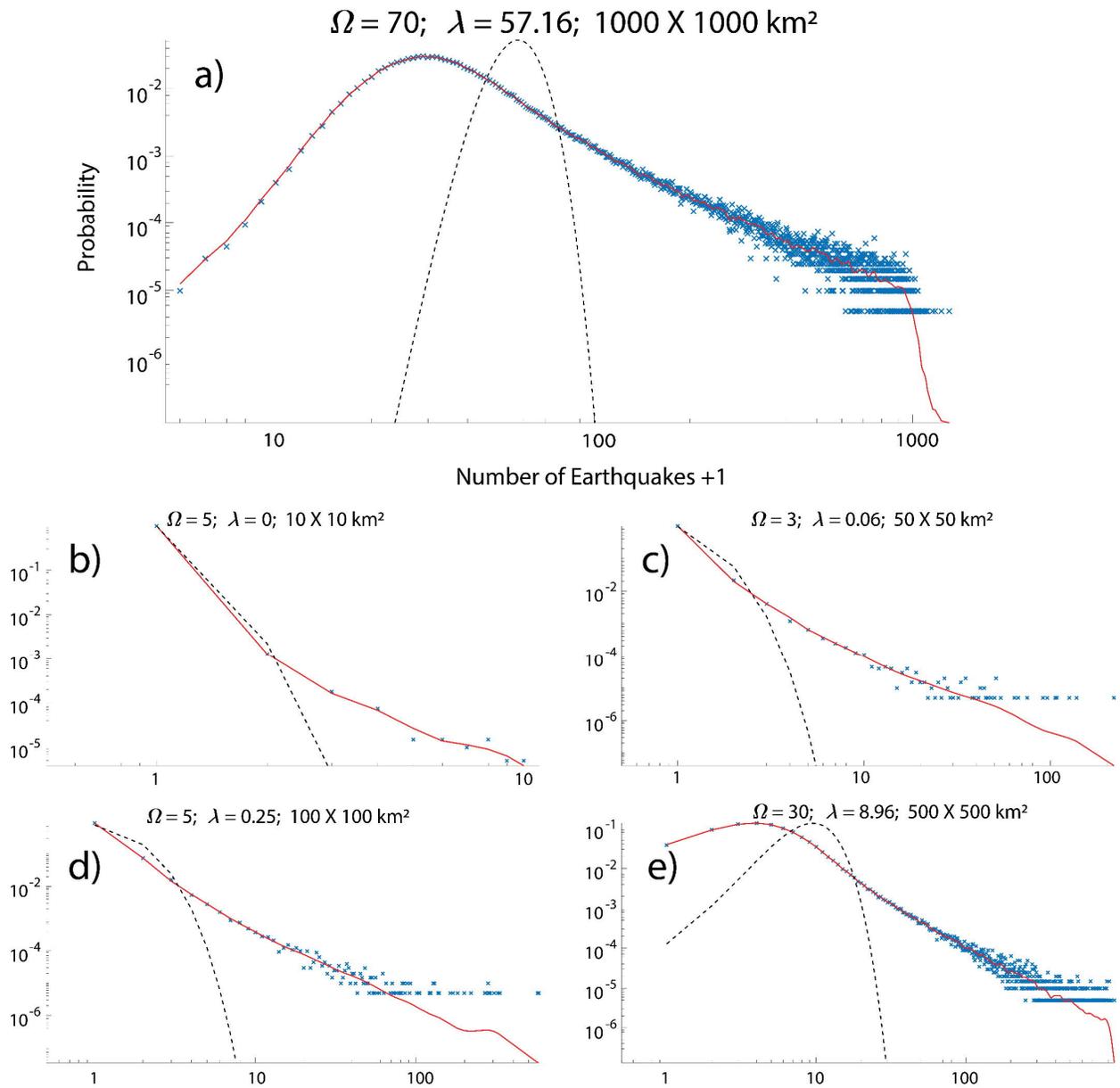

**Figure 1**: The distribution of the number of earthquakes obtained in all the simulations in one of the randomly chosen spatial bin is shown using blue crosses; these simulations correspond to pseudo-prospective experiment with the synthetic data for the testing period (30,35) years with minimum magnitude of testing catalogue being 5; Red curve shows the smoothed estimate of the empirical distribution with $\Omega$ parameter shown at the top of the panels, which has been obtained using the algorithm proposed in Supplementary Text S3; The black dashed curves show the best possible fits to the empirical distributions if the distributions of the number of earthquakes observed in a simulation are assumed to be Poissonian with rate $\lambda$ being equal to the average number of earthquakes observed in all the simulations.

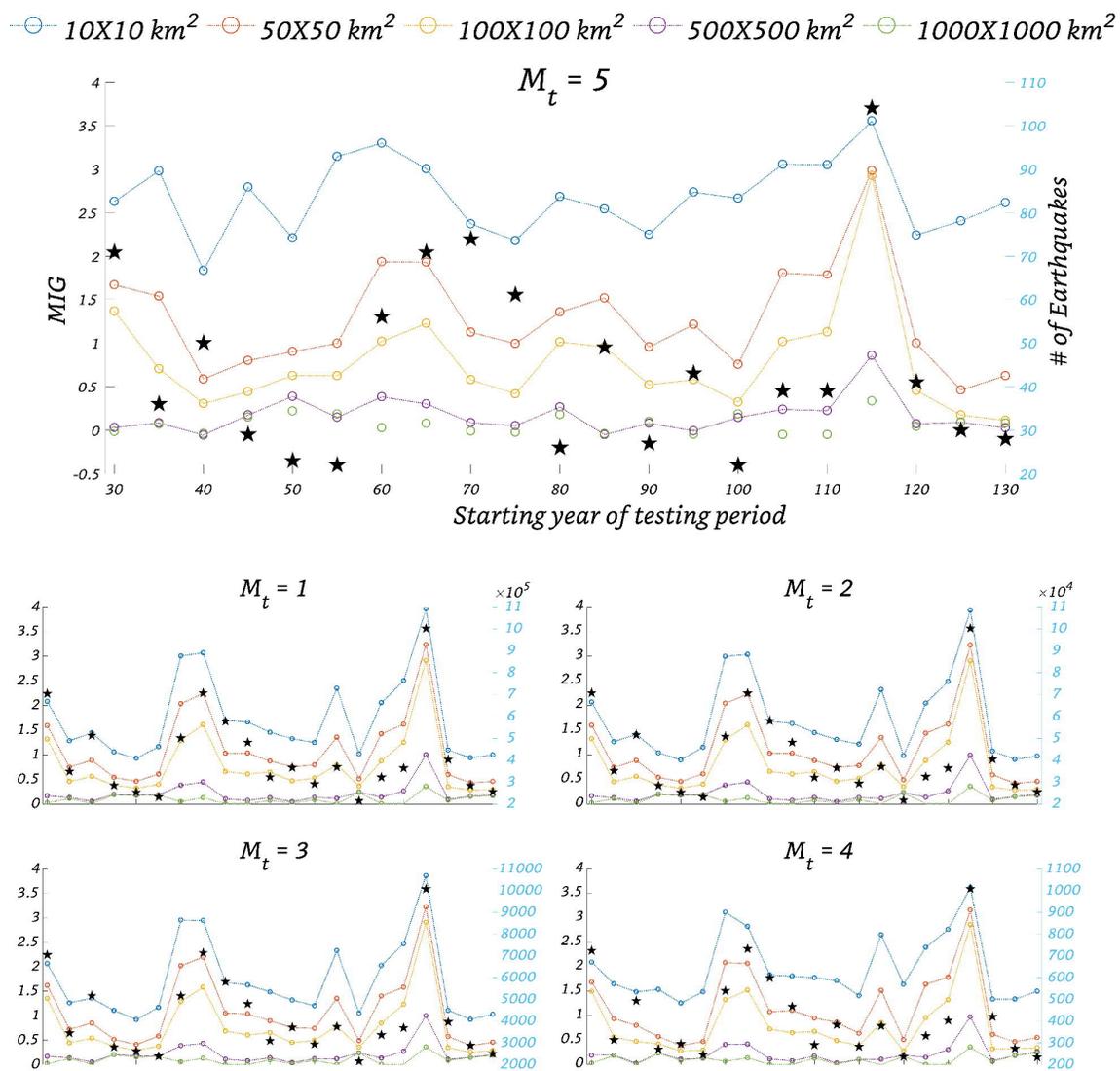

Figure 2: Mean Information Gain (MIG) defined in expression (4) of Model 1 over Model 2 for the 20 five-year-long testing periods for 5 different magnitude thresholds of the testing catalog, which are shown at the top of each of the panels; The different coloured circles correspond to the five spatial resolution at  which the experiments were conducted in each testing period; Black stars show the number of earthquakes larger than the magnitude threshold that were observed during the testing period; The starting year of the testing period is indicated on the x-axis of the top panel and is common to all the panels.The positive values of MIG confirm the superiority of Model 1 over Model 2 in all cases.

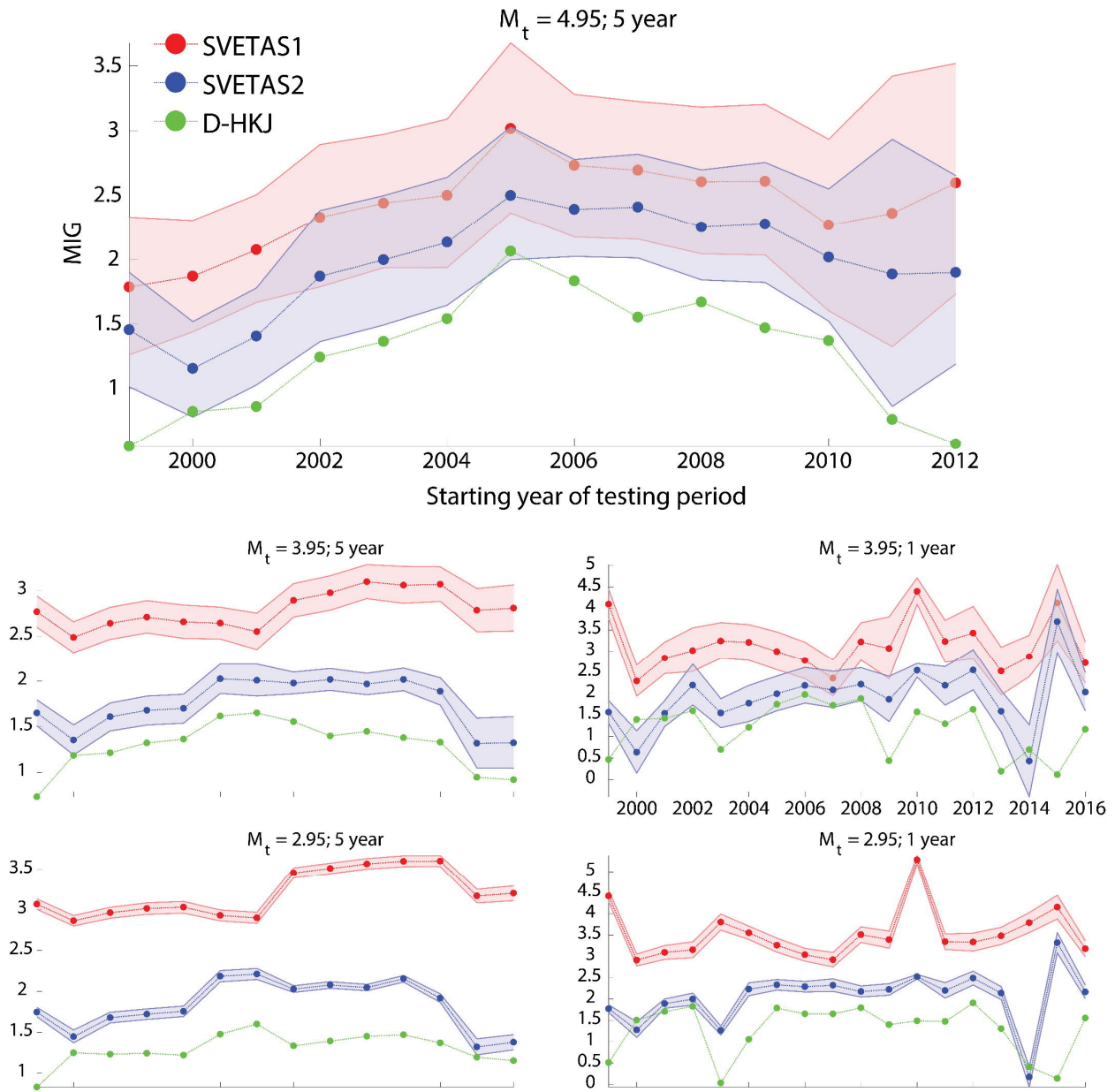

Figure 3: Time series of mean information gain (MIG) of the SVETAS1 (red), SVETAS2 (blue) and D-HKJ (green) over the spatially and temporally homogeneous Poisson process (STHPP) respectively in 5 pseudo-prospective experiments with different duration of testing period and minimum magnitude threshold of the testing catalog, which are indicated at the top of each panel; The x-axis shows the starting year of each testing period and is common among the experiments with the same duration of the testing period.

# FORECASTING THE FULL DISTRIBUTION OF EARTHQUAKE

# NUMBERS IS FAIR, ROBUST AND BETTER

Shyam Nandan[1], Guy Ouillon[2], Didier Sornette[3], Stefan Wiemer[1]

## This file contains

3 text sections (Text S1-S3) describing technical procedures and 4 figures (Figure S1-S4)

## S1 Inverting the spatially variable parameters of the SVETAS model

1. We first assume that the number of spatial Voronoi partitions (i.e. cells) that are necessary to capture the spatial variation of the parameters $\{\mu, K, a, c, \omega, \tau\}$ of the ETAS model described in Equation 2 (Section 2) is q. In each of these q partitions, $S = \{S_1, S_2, S_3, \dots, S_q\}$, the parameters $\{\mu, K, a, c, \omega, \tau\}$ are assumed to be piecewise constant. So, each partition $S_{i:1 \text{ to } q}$ has its own set of constant parameters $\{\mu, K, a, c, \omega, \tau\}_{i:1 \text{ to } q}$. The parameters $\{d, \gamma, \rho\}$ are the same in all the partitions. This assumption is made for the sake of computational simplicity and stability. To divide the study region into q partitions, we draw q points without repetition from the set of known locations of earthquakes in the catalogue to be used as centres of the Voronoi partitions, while ensuring that within each of the resulting spatial partitions there are at least 100 earthquakes. This limit of 100 earthquakes is imposed to ensure that estimated parameters of the ETAS model are stable. Note that drawing the Voronoi centres from locations of earthquakes in the catalogue, rather than randomly choosing them within the study region, allows for a higher resolution in areas that are densely populated by seismicity. The modified sampling technique is thus more natural and data-driven.

2. We then use the extended EM algorithm proposed by us [Nandan et al., 2017] to invert for the parameters $\{\mu, K, a, c, \omega, \tau\}$ is each of the spatial partitions along with the global estimates of the parameters $\{d, \rho, \gamma\}$.

3. We then compute the penalised log-likelihood score for the model, $BIC = -2LL + N_{pars} \log N$, where $N_{pars} = 8q + 3$, as each spatial partition is characterised by one parameter from the background seismicity rate, two productivity parameters, three Omori parameters and two Voronoi centre parameters, along with three global parameters corresponding to the spatial kernel. Note that LL and N respectively correspond to the expected maximum log-likelihood score obtained upon the calibration of the ETAS model using the EM algorithm and to the number of earthquakes used for the calibration of the ETAS model.

4. We then repeat the Steps 1-3 1000 times with different realisations of q Voronoi centres selected randomly from the list of the earthquakes and store the estimate parameters and BIC.

5. We then repeat steps 1 to 4 with a value of q increasing from 1 to 150.

Having calibrated 150,000 models with varying degrees of complexity, we then rank these models according to their BIC and select the top 1% solution for computing an ensemble model, while giving the $i^{th}$ model out of the M selected models the weight of $\frac{\exp\left[-\frac{BIC_i}{N}\right]}{\sum_{i=1}^{M}\exp\left[-\frac{BIC_i}{N}\right]}$.

On calibration of the SVETAS model on the primary earthquake catalogue (shown in Figure 1) covering the entire period (1981-2017), we find the ensemble estimates of the global parameters: $d = 0.20 \text{ km}^2, \gamma = 1.21, \rho = 0.58$. In Figure S3, we show the spatial variation of the six other ETAS parameters. We replace the spatially variable estimate of the parameter K by the branching ratio (n), which is a more informative parameter. It quantifies the average number of direct aftershocks triggered by any earthquake averaged over all sizes.

Note that the above procedure of calibration of the SVETAS model is repeated for all the training catalogues such that experiments remain prospective.

## S2 Simulating an ETAS catalogue for a testing period

Having calibrated the ETAS models on the training catalogue, the seismicity forecast for the corresponding testing period for the two philosophically different variants of the ETAS models can be constructed by adding the contribution of the two components: Type 1 and Type 2 earthquakes. Note that the first variant forecasts the full distribution of the number of earthquakes in each of the spatial bins. The second variant only forecasts the expected number of earthquakes in each of the spatial bin. This model further assumes that the distribution of the number of earthquakes in each of the bins is Poissonian. So, to construct the forecast of the first variant, we need to perform many simulations (~ 5,000,000) and count the total number of earthquakes (including both types) in each of them, thereby creating an empirical distribution in each of the bin. This empirical distribution is further smoothed using the method proposed in Supplementary Text S3 to obtain the forecast in each bin. On the other hand, the forecast for the second variant is obtained by estimating the average number of earthquakes observed in all the simulations in a given spatial bin.

## S2.1 Type 1 earthquakes

These earthquakes include the direct descendants of the earthquakes in the training catalogue as well as the cascade of earthquakes that would be activated by those direct aftershocks. To obtain the location time and magnitude of these earthquakes we use the following simulation scheme:

1. We first simulate the first-generation aftershocks for all the earthquakes in the training period. An earthquake with magnitude $m_i$ can trigger on average $K_i e^{[a_i(m_i - M_c)]}$ aftershocks with magnitude larger than $M_c$, where $M_c$ is the magnitude of completeness. $K_i$ and $a_i$ represent the ensemble estimate of the productivity parameters at the location of the $i^{th}$ earthquake.

   The times of those aftershocks must be simulated from the time kernel $(t - t_i + c)^{-1-\omega} \exp(-\frac{t-t_i}{\tau})$. However, this time kernel does not allow us to analytically simulate the times of direct aftershocks, which is of extreme importance as we want to perform numerous simulations. As a result, we make the following approximation. We assume that time kernel can be expressed piecewise using the following equation:

$$\psi(t) = \begin{array}{ll} C_1(t - t_i + c)^{-1-\omega} & \forall\, t < \tau \\ C_2 \exp\left(-\dfrac{t - t_i}{\tau}\right) & \forall\, t \geq \tau \end{array} \tag{S1}$$

The constants $C_1$ and $C_2$ are such that $C_1(\tau - t_i + c)^{-1-\omega} = C_2 \exp\left(-\frac{\tau - t_i}{\tau}\right)$ and

$\int_0^\tau C_1(t - t_i + c)^{-1-\omega} \, dt + \int_\tau^\infty C_2 \exp\left(-\frac{t - t_i}{\tau}\right) dt = 1$. In order to simulate the times of the

direct aftershocks of the $i^{th}$ event with this approximation, we first estimate the fraction of

aftershocks that correspond to the power law part ($f_1$) and the exponential tail ($f_2$) of the

PDF respectively. Note that, $f_1 + f_2 = 1$. We then draw a uniform random number

between 0 and 1. If the number falls between the 0 and $f_1$, the aftershock is assigned to the

power law part of the PDF otherwise it belongs to the exponential tail of the PDF. The time

of the aftershock which belongs to the power-law part of the PDF is given by:

$$t_{ji} = \left[\frac{1}{c_i^{\omega_i}} - U_j(0,1)\left(\frac{1}{c_i^{\omega_i}} - \frac{1}{(c_i + \tau_i)^{\omega_i}}\right)\right]^{-\frac{1}{\omega_i}} - c_i + t_i \qquad (S2)$$

If the aftershock belongs to the exponential tail, its time can be simulated using:

$$t_{ji} = -\tau_i \ln\left(1 - U_j(0,1)\right) + t_i \qquad (S3)$$

In the above equations, $t_{ji}$ is the time of the $j^{th}$ aftershock that has been triggered by the $i^{th}$

earthquake in the training catalogue, which is counted from the time of the $i^{th}$

earthquake; $U_j(0,1)$ is a random number drawn uniformly in the interval [0,1]; $c_i$, $\tau_i$ and $\omega_i$

are the ensemble estimate of the parameters of the time kernel at the location of $i^{th}$

earthquake that occurred at time $t_i$.

Similar to the time of the aftershocks, we simulate the Cartesian coordinates of the direct

aftershocks $\{x_{ji}, y_{ji}\}$ of the mainshock $\{m_i, x_i, y_i\}$ using the following equations:

$$r_{ji} = \left[d \exp\{\gamma(m_i - M_c)\}\left\{(1 - U_j^2)^{-\frac{1}{\rho}} - 1\right\}\right]^{\frac{1}{2}}, \qquad \theta_j = 2\pi U_j^3 \qquad (S4)$$

$$x_{ji} = r_{ji}\cos\theta_j, \qquad y_{ji} = r_{ji}\sin\theta_j$$

In the above equation, $r_{ji}$ and $\theta_j$ are respectively the simulated distance and the azimuth of the $j^{th}$ aftershock from its mainshock with index $i$; $U_j^2$ and $U_j^3$ are random numbers drawn uniformly in the interval [0,1]; $d$, $\gamma$ and $\rho$ are the ensemble estimates of the spatially invariant parameters of the spatial kernel.

We then simulate the magnitude of the direct aftershocks using the following equation:

$$m_j = \frac{-\log\left(1 - U_j^4\right)}{\beta} + M_c \qquad (S5)$$

In the above equation, $U_j^4$ is a number drawn uniformly at random from the interval [0,1] and $\beta = \log(10)$ is the assumed global exponent of the GR law.

2. Having simulated the direct aftershocks of all the earthquakes of the training catalogue, we then reject all those direct aftershocks that do not fall in the testing period.

3. All the direct aftershocks selected in step 2 are then treated as new mainshocks that are then allowed to trigger their direct aftershocks whose time, location and magnitude are simulated as in step 1. Again, the direct aftershock selection criteria outlined in step 2 is applied.

4. Steps 2 and 3 are repeated until no more aftershocks are simulated.

5. We then count the number of earthquakes with magnitude larger than the minimum magnitude threshold of the testing catalog in each of the spatial bins which constitute the testing region and store it.

6. We repeat steps 1 to 5 are repeated $5 \times 10^6$ times.

## Type 2 earthquakes

These earthquakes include the background earthquakes that are expected to occur during the testing period as well as the cascade of aftershocks that are initiated by those background earthquakes. To account for the contribution of these earthquakes during the testing period, we first simulate the background seismicity. To simulate the background earthquakes, we use the key idea that the locations of past background earthquakes reveal the probable location of future background earthquakes. In practice, we implement this idea by first drawing $U_i$, a number uniformly drawn at random from the interval $\left[0, \frac{T_{train}}{T_{test}}\right]$ and comparing it to the independence probability $IP_i$ of the $i^{th}$ earthquake in the training catalogue. Note that, $IP_i$ quantifies the probability that the $i^{th}$ earthquake in the training catalogue has not been triggered by a previous one. If $U_i \leq IP_i$, then the location of the $i^{th}$ earthquake is selected as the location of a new background event. By generating the random number between $\left[0, \frac{T_{train}}{T_{test}}\right]$, we ensure that the expected number of background earthquakes during the testing period is $\frac{T_{test}}{T_{train}} \sum_i IP_i$. This process is repeated sequentially for all the earthquakes in the training catalogue. For the selected seed events, we discard their actual time and magnitude information and only retain their location information. However, we perturb the location of the $i^{th}$ background earthquake using a Gaussian kernel with bandwidth $\sigma_i$, where $\sigma_i$ is taken as the distance of the nearest neighbour background earthquake from the $i^{th}$ earthquake. This perturbation mimics the nearest neighbor smoothing that is commonly used to construct the smoothed seismicity forecast. We then assign times to these events that are distributed uniformly randomly between $[T_{train}, T_{train} + T_{test}]$ and the magnitudes of these events are simulated as described in the algorithm for the simulation of the Type 1 earthquakes. We then simulate the cascade of aftershocks corresponding to all the simulated seed events as described for Type 1 earthquakes and store the total number of earthquakes with magnitude larger than the minimum magnitude threshold of the testing catalog in each of the spatial bins.

## S3 Smoothing using Gaussian kernels with adaptive bandwidth

Assume that in a given spatial bin we have $n_i$ distinct values of number of earthquakes that occurred in a simulation. Let us also assume that the number of simulations in which $n_i$ is observed is $m_i$, where $\sum_i m_i$ is equal to total number of simulations that have been conducted. We can then obtain the smoothed estimate of the probability ($\Pr(n)$) for any arbitrary value of $n \in \mathbb{N} \cup \{0\}$ using:

$$\Pr(n) = \frac{1}{\sum_i m_i} \sum_i \frac{m_i \int_{n-0.5}^{n+0.5} K(n'|n_i, \sigma_i) dn'}{\int_{-0.5}^{\infty} K(n'|n_i, \sigma_i) dn'} \tag{S6}$$

Where, $K(n'|n_i, \sigma_i) = \frac{1}{\sqrt{2\pi\sigma_i^2}} \exp\left[-\frac{1}{2}\left(\frac{n'-n_i}{\sigma_i}\right)^2\right]$ is Gaussian kernel with bandwidth $\sigma_i$. $\sigma_i$ is the expected distance to the nearest neighbor scaled by a factor $\Omega$, or $\sigma_i = \frac{\sum_j (n_i - n_j) m_j}{\sum_j m_j} * \frac{1}{\Omega}$, where $\Omega > 0$. The normalization by $\int_{-0.5}^{\infty} K(n'|n_i, \sigma_i) dn'$ ensures that the kernel integrates to 1 with in the discrete support range $[0, \infty]$, which is equivalent to the range $[-0.5, \infty]$ in the continuous space. Furthermore, the integration $\int_{n-0.5}^{n+0.5} K(n'|n_i, \sigma_i) dn'$ ensures the discretization of the continuous kernel for the discrete values that n could assume.

This smoothing scheme allows us to obtain the non-zero value of $Pr$ for all values of n. The only free parameter in the proposed smoothing scheme is $\Omega$, which determines, to some extent, the amount of smoothing for the $i^{th}$ bin. In order to optimize the value of $\Omega$, which is not known a priory, we use the following scheme:

1. We take the numbers of earthquakes observed in a given spatial bin during all the simulations and divide it into training and validation set. The training set consists of 90% of the data and the validation set consists of 10% of the data.

2. Using the observed numbers and their frequencies in the training data we compute the probabilities of the numbers in the validation data. The total log-likelihood computed on the validation data for an arbitrary value of $\Omega$ is given by:

$$LL(\Omega) = \sum_j m_j^v \left[ \ln \left\{ \sum_i \frac{m_i \int_{n_j^v-0.5}^{n_j^v+0.5} K\left(n' \middle| n_i^t, \frac{n_i^t - n_{i-1}^t}{\Omega}\right) dn'}{\int_{-0.5}^{\infty} K\left(n' \middle| n_i, \frac{n_i^t - n_{i-1}^t}{\Omega}\right) dn'} \right\} - \ln \sum_i m_i^t \right] \quad (S7)$$

where, the superscripts $v$ and $t$ represent the training and the validation set respectively. All other symbols have the same meaning as in Equation S6.

3. We obtain the maximum likelihood estimate of $\widehat{\Omega}$ and the maximum log-likelihood $MLL$ by maximizing $LL(\Omega)$ with respect to $\Omega$.

4. We repeat this step 1 and 3 several (100) times by permuting the data in the training and validation set and store corresponding values of $\widehat{\Omega}_k$ and $MLL_k$ for the $k^{th}$ iteration.

5. Finally, we obtain the ensemble estimate of $\widehat{\Omega}$ using:

$$\widehat{\Omega} = \sum_k \frac{\exp\left(\frac{MLL_k}{\sum_j m_j^v}\right) \widehat{\Omega}_k}{\sum_k \exp\left(\frac{MLL_k}{\sum_j m_j^v}\right) \widehat{\Omega}_k} \quad (S8)$$

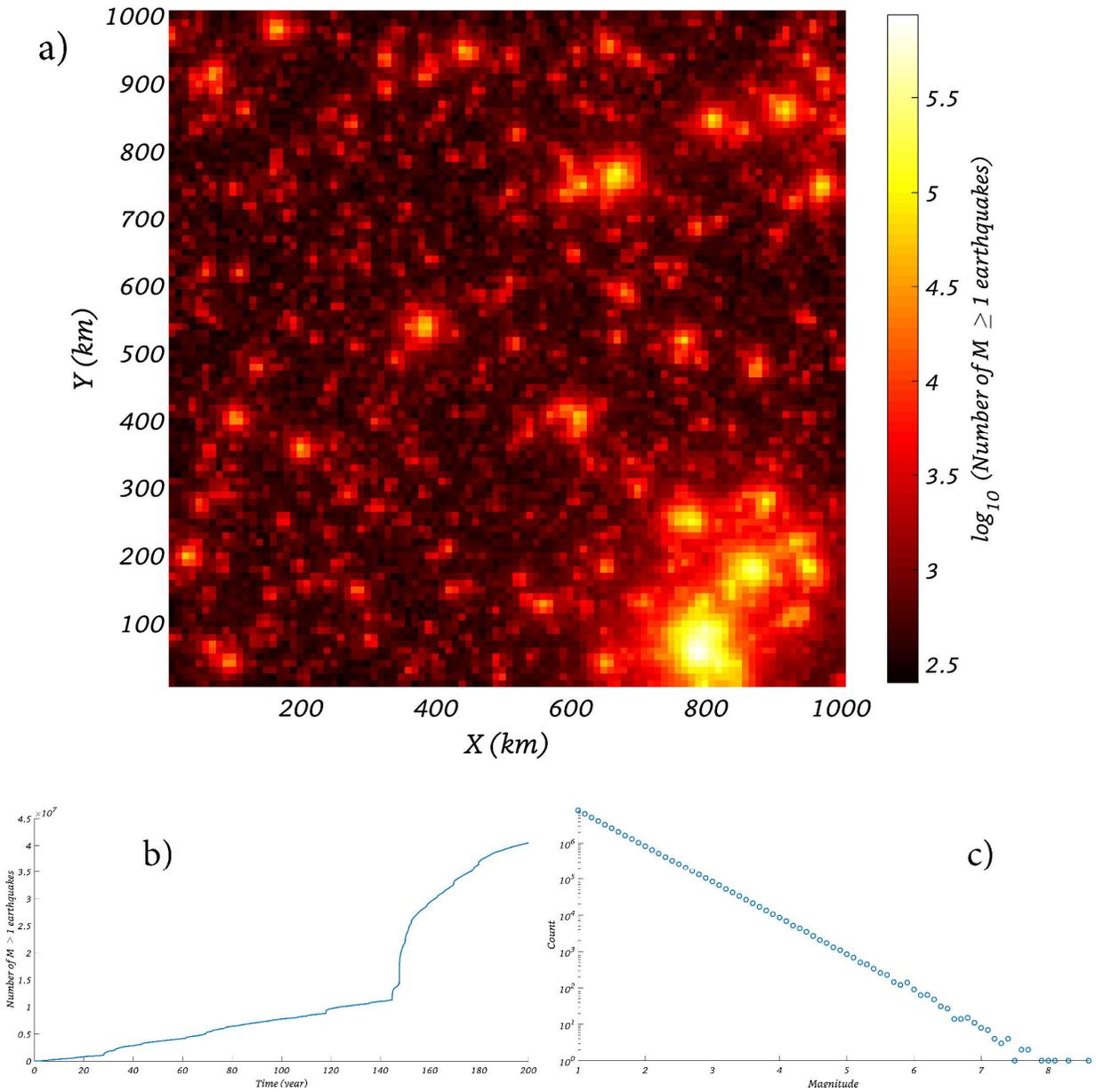

Figure S4: (a) Spatial density of simulated catalog with the parameter setting specified in section 3.1; (b) Evolution of total number of earthquakes in the simulated catalog as a function of time; (c) Frequency-Magnitude distribution of the simulated catalog.

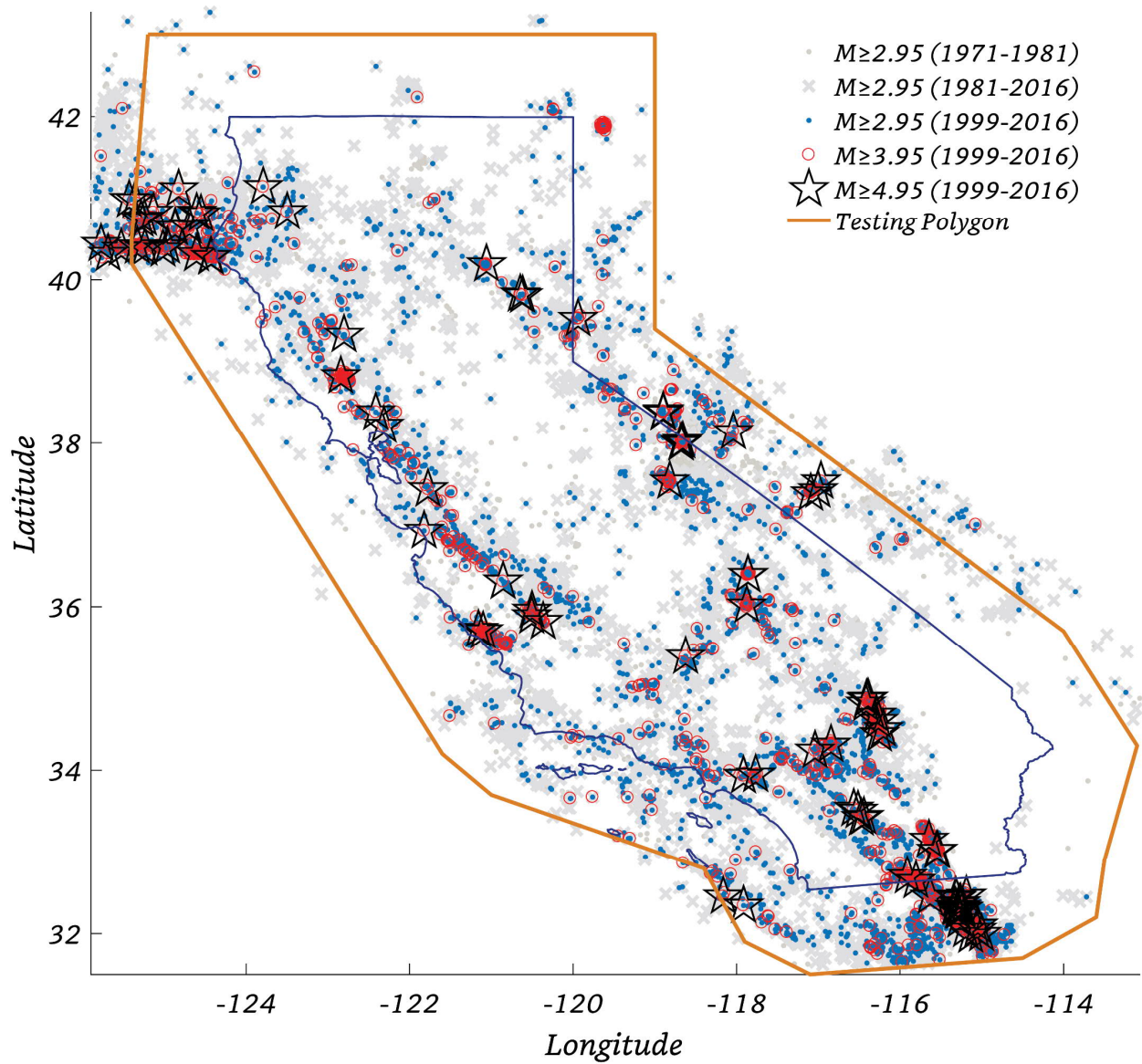

Figure S5: Californian catalog used in this study: grey dots show the locations of auxiliary earthquakes ($M \geq 2.95$; $1971 - 1981$); grey crosses show the locations of all primary earthquakes ($M \geq 2.95$; $1981 - 2016$); blue dots, red circles and black stars show the locations of $M \geq 2.95$, $M \geq 3.95$ and $M \geq 4.95$ earthquakes that occurred during the entire testing period (1999-2016); only earthquakes within the testing polygon are involved in testing.

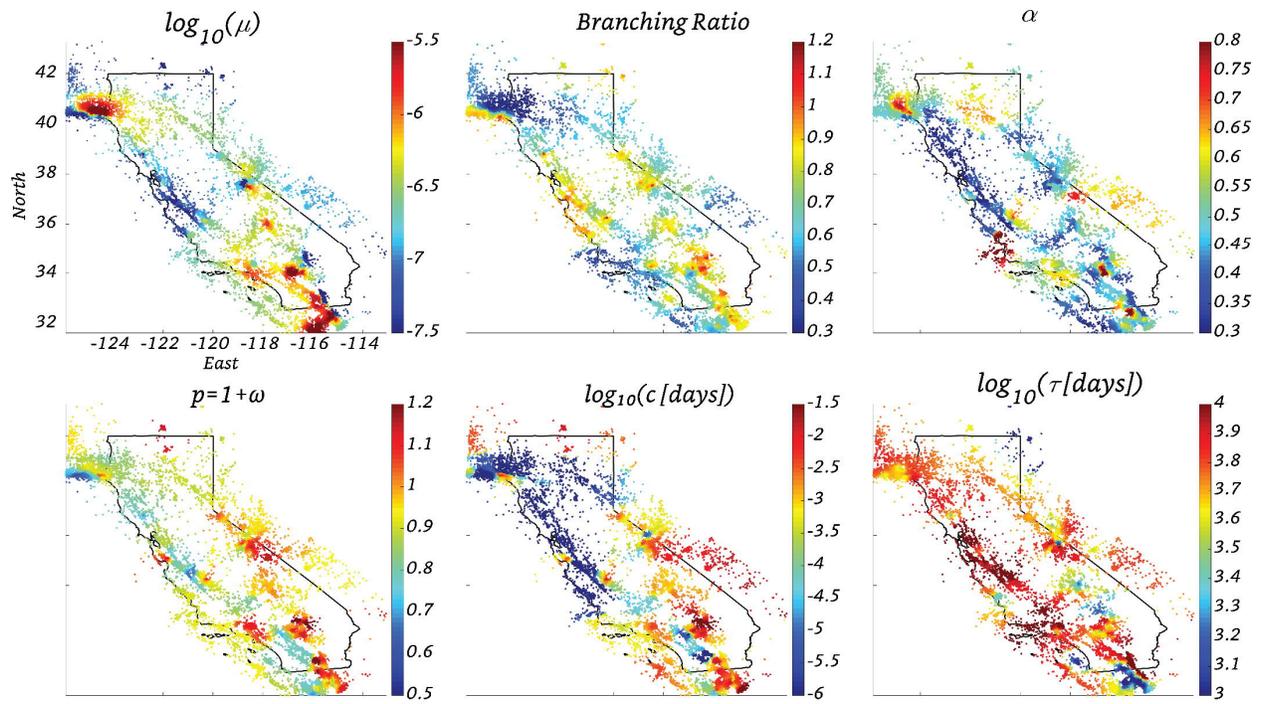

Figure 1: Spatial variation of the ETAS parameters: background seismicity rate ($\mu$), productivity parameters: branching ratio and $\alpha(= \frac{a}{log\,g(10)})$ and Omori parameters: $c$ and $p(= 1 + \omega)$ and $\tau$.

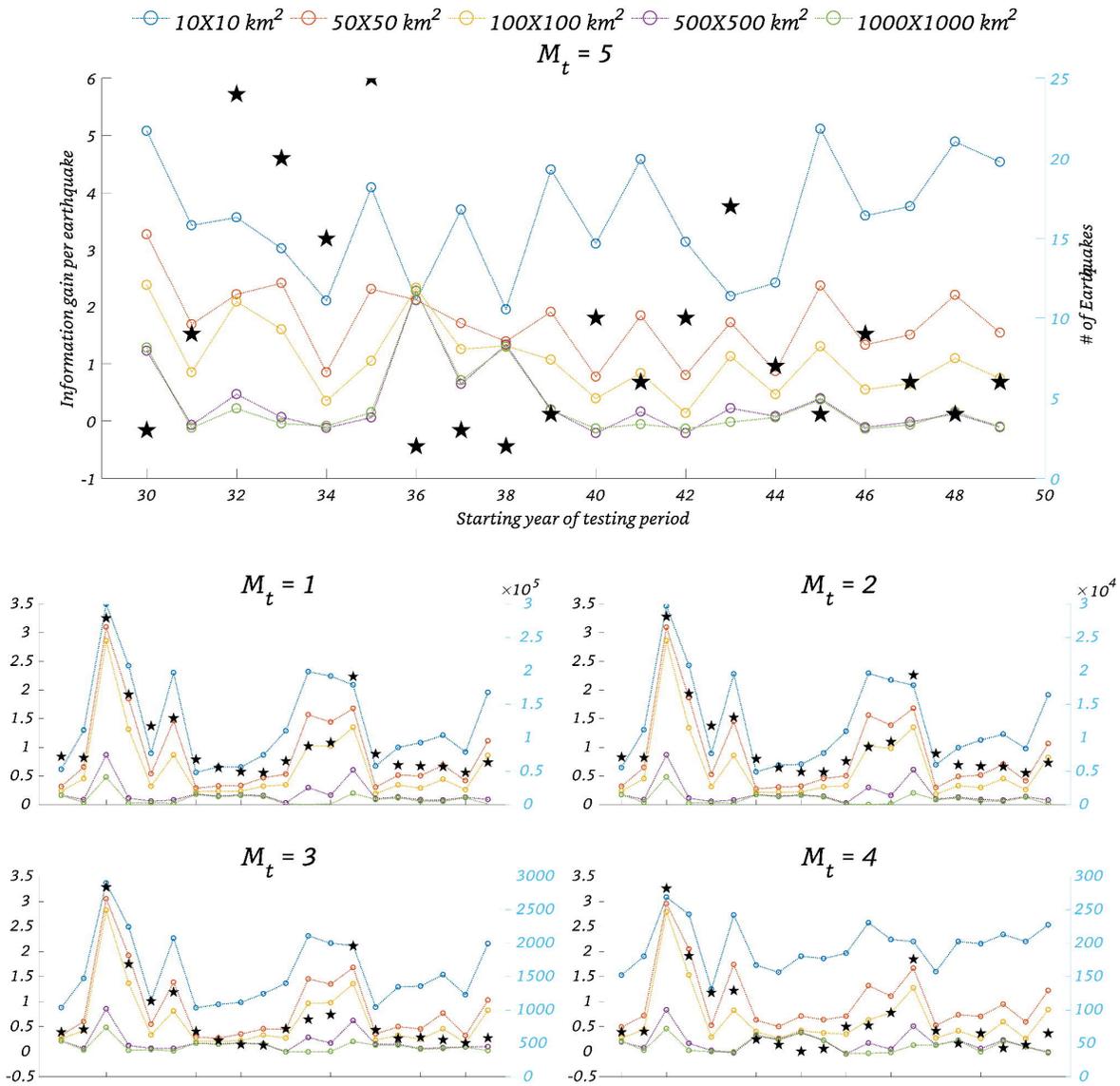

Figure S6: Mean Information Gain (MIG) of the Model 1 over Model 2 for the 20 one-year-long testing periods for 5 different magnitude thresholds of the testing catalog, which are shown at the top of each of the panels; The different coloured circles correspond to the five spatial resolution at which the experiments were conducted in each testing period; Black stars show the number of earthquakes larger than the magnitude threshold that were observed during the testing period; The starting year of the testing period is indicated on the x-axis of the top panel and is common to all the panels.